\documentclass[aps,prd,twocolumn,eqsecnum,showpacs,amsmath,nofootinbib]{revtex4-1}

\usepackage{color,graphicx}
\usepackage{amsfonts}
\usepackage{amssymb}
\begin{document}

%<<<<<<<<<<<<< TITLE >>>>>>>>>>>>>>>%
\title{
High energy collision of two particles in wormhole spacetimes
}

%<<<<<<<<<<<<< AUTHOR >>>>>>>>>>>>>>>%   
\author{Naoki Tsukamoto} %\email{tsukamoto@fudan.edu.cn}
\author{Cosimo Bambi} 
\email[Corresponding author: ]{bambi@fudan.edu.cn}

\affiliation{
Center for Field Theory and Particle Physics and Department of Physics, Fudan University, 200433 Shanghai, China 
}
%<<<<<<<<<<<<< DATE >>>>>>>>>>>>>>>%
\date{\today}

%======================================%
%<<<<<<<<<<<<< ABSTRACT >>>>>>>>>>>>>>>%
%======================================%
\begin{abstract}
We study the collision of two particles in the Teo wormhole spacetime, in which the wormhole is stationary and axisymmetric. We show that a non-rotating Teo wormhole cannot be a particle accelerator, while a rotating Teo wormhole can be used to accelerate particles and create high energy collisions because of the deep effective potential of the colliding particles. The process is different from that in the vicinity of a near-extremal black hole, since here there is no event horizon. This is the first example of particle collision with high center-of-mass energy in a spacetime with no event horizon, no naked singularity, and not being extremal in a clear sense. The process can unlikely have direct implications for astrophysical observations, but it is interesting as a tool to investigate wormhole instabilities.
\end{abstract}

%<<<<<<<<<<<<< PACS NUMBER >>>>>>>>>>>>>>>%

%<<<<<<<<<<<<< preprint number >>>>>>>>>>>>>>>%
%\preprint{
%}

\maketitle

%\tableofcontents

\section{Introduction}
Recently, Ba\~{n}ados, Silk and West have rediscovered that the center-of-mass 
energy of a collision of two particles that are at rest at infinity can be arbitrary high 
near the horizon of an extremal Kerr black hole if the particle angular momentum 
is fine-tuned to a critical value~\cite{Banados:2009pr}.
The phenomenon is now called Ba\~{n}ados-Silk-West (BSW) effect, but actually it 
was found for the first time by Piran, Shaham and Katz in 1975 in a sequence of papers 
on the collisional 
Penrose process~\cite{Piran_Shaham_Katz_1975,Piran_Shaham,Piran:1977dm}.
The properties of high energy collisions in the vicinity of extremal and almost 
extremal Kerr black holes have been investigated in detail~\cite{Grib:2010dz}. 
For a review, see e.g. Harada and Kimura~\cite{Harada:2014vka}. 
The BSW process has been criticized in~\cite{Jacobson:2009zg,Berti:2009bk}. First, 
the proper time of a critical particle to reach the event horizon of an extremal 
Kerr black hole is infinite. 
Second, it is widely believed that astrophysical black holes cannot be extremal. 
The Thorne bound for the spin parameter of a Kerr black hole accreting from 
a thin disk is $J/M^2 \le 0.998$~\cite{Thorne:1974ve}, even if
it can be violated in the cases of super-Eddington 
accretion~\cite{Abramowicz_Jaroszynski_Sikora_1978,Abramowicz_Lasota_1980} 
and of deviations from the Kerr background~\cite{super}.

From the relativistic point of view, high energy collisions of particles on curved 
spacetimes are interesting because the phenomenon would have close 
relationships with other relativistic effects, such as the emission of gravitational 
waves~\cite{Harada:2011pg}, the self-gravity of colliding objects~\cite{Kimura:2010qy}, 
and instabilities of the background 
spacetime~\cite{Lake:2010bq,Kimura:2010qy,Tsukamoto:2013dna}. 
In astrophysical situations, the true upper bound of the center-of-mass energy for 
a particle collision would be affected by these relativistic effects.

The electromagnetic 
counterpart of the BSW effect in the extremal Reissner-Nordstr\"om black hole 
spacetime was found by Zaslavskii~\cite{Zaslavskii:2010aw},
although the extremal Reissner-Nordstr\"om black hole can unlikely be considered an astrophysical object.
The details of the BSW collision in the Reissner-Nordstr\"om spacetime have been investigated 
to understand the BSW effect deeply~\cite{Kimura:2010qy,Nemoto:2012cq,Patil:2011uf}
since the Reissner-Nordstr\"om spacetime has a simpler structure than the Kerr spacetime.

The BSW effect has been also studied in a variety of near-extremal black hole 
spacetimes~\cite{Zaslavskii:2010jd, Kimura:2010qy}, including cases in higher 
dimensions~\cite{Abdujabbarov:2013qka, Tsukamoto:2013dna}, and some 
near-extremal non-black-hole spacetimes~\cite{Patil:2011yb, Zaslavskii:2013nra}. 
These recent studies have revealed that the BSW effect is a universal property in extremal and near-extremal spacetimes,
while particle accelerations would also occur in non-near-extremal spacetimes 
if a gravitational potential is deep~\cite{Harada:2014vka}.  
In this paper, we will show clearly that particle collisions with a high center-of-mass energy can occur because of a deep effective potential.

The BSW effect can be used to study the instability of the background spacetime. Kimura~\textit{et al.} pointed out that a BSW collision may generate a new black hole in the vicinity of extremal or almost extremal black holes~\cite{Kimura:2010qy}. This implies that extremal black holes are unstable against the BSW process. Tsukamoto~\textit{et al.}~\cite{Tsukamoto:2013dna} showed that there are analogies between the instability of extremal horizons suggested by the BSW effect 
and the test-field instability of extremal horizons studied by Aretakis~\cite{Aretakis:2011ha}.

General relativity permits gravitating objects with a non-trivial topology such as wormholes (for a review on wormhole spacetimes, see e.g. Visser~\cite{Visser_1995}). Astrophysical observations give some constraints~\cite{newref1}, but they cannot exclude the existence of these objects in our Universe. For instance, some traversable wormholes are still viable candidates to explain the supermassive objects at the centers of galaxies~\cite{Bambi}. The Ellis wormhole is the simplest traversable wormhole of the Morris-Thorne class~\cite{Morris_Thorne_1988}. It is a static, spherically symmetric and asymptotically flat solution of Einstein's equations with a phantom scalar field~\cite{Ellis_1973}. However, the spacetime is unstable~\cite{Shinkai_Hayward_2002}, despite some early studies claiming its stability~\cite{Armendariz-Picon_2002}.

In 2013, Bronnikov \textit{et al.} considered the Ellis wormhole metric, but with the spacetime filled with a perfect fluid with negative density and a source-free radial electric or magnetic field. With this set-up, they proved that their wormhole was linearly stable under both spherical and axial perturbations~\cite{Bronnikov:2013coa}. It seems to be the first example of a stable wormhole without a thin-shell in general relativity. Their result clearly shows that the stability of a wormhole does not only depend on the metric, but also on the matter content supporting the wormhole. Interestingly, the center-of-mass energy of a collision between two particles is independent of the matter content supporting the wormhole. The possible instability suggested by the BSW effect can thus be a useful tool to study the instability of the spacetime.

In this paper, we will investigate the collision of two particles in the Teo wormhole spacetime. Such a metric was proposed by Teo in~\cite{Teo:1998dp} and describes a simple rotating wormhole. Because of the absence of any event horizon, the process is different from the BSW effect. We find that the spin of the wormhole dramatically changes the picture. Non-rotating Teo wormholes cannot be accelerators in a particle collisional scenario, while rotating Teo wormholes can potentially accelerate particles and create collisions with a high center-of-mass energy.

The paper is organized as follows.
In Sec.~II, we consider the motion of a test particle in the Teo wormhole spacetime. 
In Sec.~III, we investigate the collision of two particles in the vicinity of a Teo wormhole.
In Sec.~IV, we discuss and summarize our results.
In appendix~A, we consider particle collisions in the Ellis wormhole spacetime.
In appendix~B, we study the relation between the scalar curvature and the collision of two particles with a high center-of-mass energy in wormhole spacetimes.
Throughout the paper, we use geometrical units in which the light speed $c$ and Newton's constant $G_{\rm N}$ are 1, unless stated otherwise.

\section{Particle Motion in the Teo wormhole Spacetime}

In this section, we will investigate the motion of a test-particle in the Teo wormhole spacetime, which is a simple example of a rotating wormhole and was suggested by Teo in~\cite{Teo:1998dp}. The line element in spherical polar coordinates is given by 
\begin{eqnarray}\label{eq:Metric}
ds^{2}
&=&-N^{2}dt^{2} +\frac{1}{1-\frac{b}{r}}dr^{2} \nonumber\\
&&+r^{2}K^{2} \left[ d\theta^{2}+\sin^{2}\theta (d\phi-\omega dt)^{2} \right] , 
\end{eqnarray}
where 
\begin{eqnarray}\label{eq:N}
N
&=&K
=1+\frac{16a^{2}d\cos^{2}\theta}{r} ,\\
\omega
&=&\frac{2a}{r^{3}} ,
\end{eqnarray}
$b$ and $d$ are positive constants~\footnote{
Here, we have introduced a positive parameter $d$ to tune the dimension of the second term on the right-hand side in Eq.~(\ref{eq:N}), because we use the geometrical units in which $c=G_{\rm N}=1$, while Teo used units in which $c=G_{\rm N}=b=1$. The dimension of $d$ is $[d]=[\mathrm{m}^{-3}]=[\mathrm{kg}^{-3}]=[\mathrm{s}^{-3}]$ in geometrical units. If we set $b=d=1$, we find the same line element as in Teo's paper~\cite{Teo:1998dp}.
},
and we shall consider the range of the radial coordinate $r\geq b$.
The throat of the wormhole is at $r=b$ and the flare-out condition~\cite{Morris_Thorne_1988} is satisfied there:
\begin{eqnarray}
\frac{b-b_{,r}r}{2b^{2}}>0 .
\end{eqnarray}
Without loss of generality, we assume a non-negative spin angular momentum $a\ge 0$. The spacetime has the time translational Killing vector $t^{\mu}\partial_{\mu}=\partial_{t}$ and the axial Killing vector $\phi^{\mu}\partial_{\mu}=\partial_{\phi}$, since it is, respectively, stationary and axisymmetric.

The throat of the wormhole has a peanut-shell-like shape (see Fig.~1 in Teo's paper~~\cite{Teo:1998dp}) and the wormhole shadow was studied by Mori~\cite{Mori_2014}. In the fast-rotating wormhole case with $a>b^{2}/2$, the ergoregion exists in the range $2a \left| \sin \theta \right| > r^{2}>b^{2}$. It is near the equatorial plane ($\theta=\pi/2$), but it does not extend to the poles ($\theta=0$ and $\pi$).

Let us now consider the motion of a test-particle with the rest mass $m$ and the angular momentum $p^{\mu}=dx^{\mu}/d\lambda$, where $x^{\mu}$ is the spacetime coordinate and $\lambda$ is a parameter defined as $\lambda \equiv \tau/m$ for a massive particle with proper time $\tau$. For the sake of simplicity, we focus on the motion of a particle on the equatorial plane $\theta=\pi/2$. We introduce a new radial coordinate $\rho$ defined by
\begin{eqnarray}
\frac{d\rho}{dr} \equiv \pm \left( 1-\frac{b}{r} \right)^{-\frac{1}{2}},
\end{eqnarray}
which assumes values in the range $-\infty < \rho <\infty$, and we rewrite the line element as 
\begin{eqnarray}
ds^{2}
&=&-dt^{2} +d\rho^{2} +r^{2}(\rho)(d\phi-\omega(\rho) dt)^{2}.
\end{eqnarray}
Without loss of generality, we can assume that the wormhole throat is at $\rho=0$ and we obtain 
\begin{eqnarray}
\rho= \pm \left[ \sqrt{r(r-b)}+b\log \left( \sqrt{\frac{r}{b}}+\sqrt{\frac{r}{b}-1} \right) \right],
\end{eqnarray}
where the sign of the right-hand side is respectively $+$ or $-$ in the region on the one side or the other side of the throat. From the forward-in-time condition $dt/d\lambda \geq 0$, the following relation holds along geodesics:
\begin{eqnarray}
\mathcal{E}(\rho)\equiv E-\omega(\rho) L \geq 0,
\end{eqnarray}
where $E\equiv -p_{\mu}t^{\mu}=-p_{t}$ and $L\equiv p_{\mu}\phi^{\mu}=p_{\phi}$ are, respectively, the conserved energy and the conserved angular momentum for a massive particle.

From the condition $-m^{2}=p^{\mu}p_{\mu}$, we obtain 
\begin{eqnarray}\label{eq:energy_eq}
\frac{1}{2}\left( \frac{d\rho}{d\lambda} \right)^{2} +V_{eff}(\rho)=0,
\end{eqnarray}
where 
\begin{eqnarray}
V_{eff}(\rho)\equiv -\frac{1}{2}R(\rho),
\end{eqnarray}
and
\begin{eqnarray}
R(\rho)\equiv -m^{2}+\mathcal{E}^{2}(\rho) -\frac{L^{2}}{r^{2}(\rho)}.
\end{eqnarray}
We can rewrite Eq.~(\ref{eq:energy_eq}) as 
\begin{eqnarray}
\frac{d\rho}{d\lambda}=\sigma_{\rho}\sqrt{R(\rho)},
\end{eqnarray}
where $\sigma_{\rho}=-1$ $(+1)$ for a particle approaching (leaving) the wormhole throat at $\rho=0$ in a region $\rho \ge 0$. In the region $\rho < 0$, $\sigma_{\rho}=-1$ $(+1)$ when the particle leaves (approaches) the wormhole throat.

Motion is possible when $V_{eff}(\rho)\leq 0$, namely $R(\rho) \geq 0$. The effective potential at infinity is
\begin{eqnarray}
\lim_{\rho \rightarrow \infty} V_{eff}(\rho)= \frac{1}{2} \left( m^{2}-E^{2} \right),
\end{eqnarray}
and a particle can be at infinity if the conserved energy is larger than its rest mass, i.e. $E^{2}\ge m^{2}$. The effective potential has an extreme value at the throat $\rho=0$:  
\begin{eqnarray}
V_{eff}(0)=\frac{1}{2} \left[ m^{2}-\left( E-\frac{2aL}{b^{3}} \right)^{2} +\frac{L^{2}}{b^{2}} \right].
\end{eqnarray}
From $V_{eff}(0)\leq 0$, a particle can be at the throat $\rho=0$ either if $L \le L_{-}$ or if $L \ge L_{+}$ for any value of the conserved energy $E$ in the case $a \geq  b^{2}/2$. When $E^{2} \geq (1-4a^{2}/b^{4})m^{2}$ in the case $a < b^{2}/2$, the condition is $L_{-} \le L \le L_{+}$, where 
\begin{eqnarray}
L_{\pm}\equiv \frac{2aE \pm \sqrt{E^{2}b^{4}+m^{2} \left( 4a^{2}-b^{4} \right) }}{4a^{2}-b^{4}} b^{3}.
\end{eqnarray}

At this point, it is useful to introduce the following dimensionless parameters
\begin{eqnarray}
a_{\ast}\equiv \frac{a}{b^{2}},\; e\equiv \frac{E}{m},\; l\equiv \frac{L}{mb},\;    
r_{\ast}\equiv \frac{r}{b},\; \rho_{\ast}\equiv \frac{\rho}{b},
\end{eqnarray} 
where $\rho_{\ast}$ and $r_{\ast}$ are related by
\begin{eqnarray}
\rho_{\ast}=\pm \left[ \sqrt{r_{\ast} \left( r_{\ast}-1 \right)} +\log \left( \sqrt{r_{\ast}}+\sqrt{r_{\ast}-1} \right) \right].
\end{eqnarray} 
Using these parameters, we can rewrite the effective potential for a particle as 
\begin{eqnarray}
\frac{2V_{eff}(\rho_{\ast})}{m^{2}}
=1-\left( e-\frac{2a_{\ast}l}{r^{3}_{\ast}(\rho_{\ast})} \right)^{2} +\frac{l^{2}}{r^{2}_{\ast}(\rho_{\ast})}.
\end{eqnarray} 
We note that $V_{eff}(\rho_{\ast})/m^{2}$ is invariant under transformations that do not change the  dimensionless parameters $e$, $l$, $a_{\ast}$, $r_{\ast}$ and $\rho_{\ast}$. An example is the transformation $E\rightarrow 2E$, $L\rightarrow 2L$ and $m\rightarrow 2m$.

Fig.~1 shows some examples of the effective potential $V_{eff}(\rho)/m^{2}$ in which the particle can reach the wormhole throat at $\rho=0$. We note that the effective potential is even with respect to the radial coordinate $\rho$, namely $V_{eff}(-\rho)=V_{eff}(\rho)$.
\begin{figure}[htbp]
\begin{center}
\includegraphics[width=87mm]{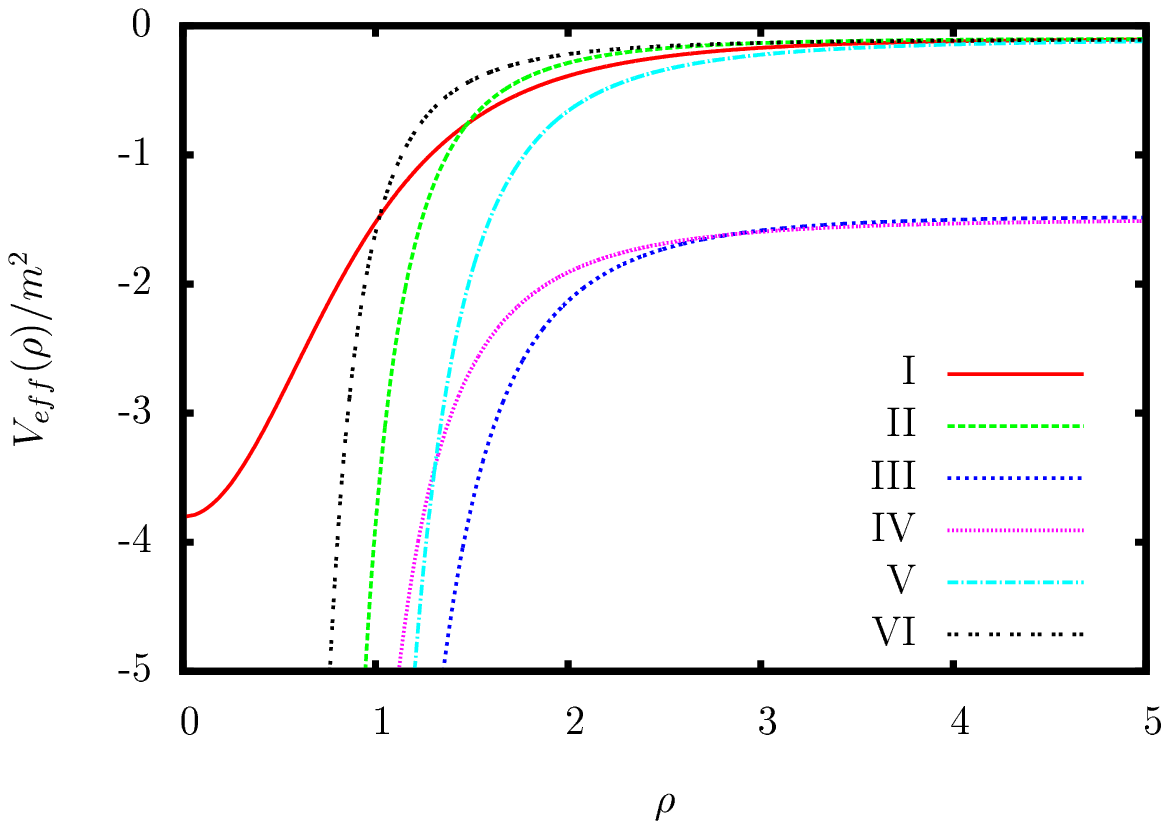}
\end{center}
\caption{Some examples of the effective potential $V_{\rm eff}(\rho)/m^{2}$ in which a particle moves from a flat region at infinity to another flat region at infinity in the Teo wormhole spacetime passing through the wormhole throat. The solid~(red), dashed~(green), dot-spaced~(blue), dotted~(magenta), dash-dotted~(cyan) and double-dot-spaced~(black) curves denote
the effective potentials of a particle with, respectively, the following set of parameters: 
I~($a=1$, $m=1$, $E=1.1$, $L=-1$, $b=1$), 
II~($a=1$, $m=1$, $E=1.1$, $L=-1$, $b=0.001$), 
III~($a=1$, $m=1$, $E=2$, $L=-2$, $b=0.001$), 
IV~($a=1$, $m=1$, $E=2$, $L=-1$, $b=0.001$),
V~($a=2$, $m=1$, $E=2.2$, $L=-1$, $b=0.001$), and 
VI~($a=1$, $m=2$, $E=2.2$, $L=-1$, $b=0.001$).
We note that the effective potentials $V_{eff}(\rho)/m^{2}$ are already quite close to their asymptotic values $1/2-E^{2}/(2m^{2})$ at $\rho=5$. We also note that the effective potentials are even with respect to $\rho$, i.e. $V_{eff}(-\rho)=V_{eff}(\rho)$, and the derivatives of the effective potential with respect to $\rho$ vanish at the throat, namely $\frac{dV_{eff}(0)}{d\rho}=0$.
}
\end{figure}

A particle that is initially at $\rho=\rho_{i}>0$ reaches the wormhole throat at $\rho=0$ in a proper time
\begin{eqnarray}
\tau =m\int^{\rho_{i}}_{0}\frac{d\rho}{\sqrt{R(\rho)}}.
\end{eqnarray} 
Fig.~1 shows that this time is finite. We remind the reader that a critical particle with a fine-tuned angular momentum reaches the event horizon in the extremal Kerr black hole spacetime in an infinite proper time.

\section{Particle Collisions in the Teo Wormhole Spacetime}

Now we want to study the collision of two particles, say particle~$(1)$ with rest mass $m_{(1)}$ and particle~$(2)$ with rest mass $m_{(2)}$. The center-of-mass energy of the collision is 
\begin{eqnarray}\label{eq:energy}
E_{CM}^{2}(\rho)
&=&-\left(p_{(1)}^{\mu}+p_{(2)}^{\mu}\right)\left(p_{(1)\mu}+p_{(2)\mu}\right)\nonumber\\
&=&m_{(1)}^{2} +m_{(2)}^{2} +2\mathcal{E}_{(1)}(\rho)\mathcal{E}_{(2)}(\rho)-\frac{2L_{(1)}L_{(2)}}{r^{2}(\rho)}  \nonumber\\
&&-2\sigma_{\rho(1)}\sigma_{\rho(2)}\sqrt{R_{(1)}(\rho)}\sqrt{R_{(2)}(\rho)},
\end{eqnarray} 
where $R_{I}(\rho)$ and $\mathcal{E}_{I}(\rho)$ are defined by
\begin{eqnarray}
R_{I}(\rho)\equiv -m_{I}^{2}+ \mathcal{E}_{I}^{2}(\rho)-\frac{L_{I}^{2}}{r^{2}(\rho)}
\end{eqnarray}
and
\begin{eqnarray}
\mathcal{E}_{I}(\rho)\equiv E_{I}-\omega(\rho)L_{I},
\end{eqnarray}
respectively, for particle $I=(1)$ and $(2)$, and where $p_{I}^{\mu}$, $L_{I}$, $\sigma_{\rho I}$, and $E_{I}$ are $p^{\mu}$, $L$, $\sigma_{\rho}$, and $E$ for $I=(1)$ and $(2)$.

We are interested in the collisions of two particles falling into the rotating wormhole from the two flat regions. With this set-up, we have $\sigma_{\rho(1)}\sigma_{\rho(2)}=-1$. The center-of-mass energy of the collision of the particles at the wormhole throat~$\rho=0$ is
\begin{eqnarray}
E_{CM}^{2}(0)
&=&m_{(1)}^{2} +m_{(2)}^{2} +2\mathcal{E}_{(1)}(0)\mathcal{E}_{(2)}(0)-\frac{2L_{(1)}L_{(2)}}{b^{2}}  \nonumber\\
&&+2\sqrt{R_{(1)}(0)}\sqrt{R_{(2)}(0)}.
\end{eqnarray} 
Here, $R_{I}(0)$ and $\mathcal{E}_{I}(0)$ are given by
\begin{eqnarray}
R_{I}(0)
= -m_{I}^{2}+ \mathcal{E}_{I}^{2}(0)-\frac{L_{I}^{2}}{b^{2}}
\end{eqnarray}
and
\begin{eqnarray}
\mathcal{E}_{I}(0)
=E_{I}-\frac{2aL_{I}}{b^{3}},
\end{eqnarray} 
respectively, for $I=(1)$ and $(2)$.

\subsection{Static and Spherically Symmetric Wormholes ($a=0$)}

If the wormhole is static and spherically symmetric, namely $a=0$, the center-of-mass energy of the head-on collision of the two particles at the wormhole throat~$\rho=0$ is
\begin{eqnarray}
E_{CM}^{2}(0)
&=&m_{(1)}^{2} +m_{(2)}^{2} +2E_{(1)}E_{(2)}-\frac{2L_{(1)}L_{(2)}}{b^{2}}  \nonumber\\
&&+2\sqrt{R_{(1)}(0)}\sqrt{R_{(2)}(0)}.
\end{eqnarray} 
If the wormhole throat is small $b\ll \left| L_{I} \right|$, the center-of-mass energy for the particle collision can be large. However, this requires that the particles have a large conserved energy $E^{2}(>m^{2}+L^{2}/b^{2})$ to reach the wormhole throat. The conclusion is that the static and spherical symmetric Teo wormhole cannot be used as particle accelerator in the particle collisional scenario.

\subsection{Small and Fast-Rotating Wormholes}

If the wormhole is small in size and it is fast-rotating, namely $b\ll a^{1/2}$, $\left| L_{I}/E_{I} \right|$, and $\left| L_{I}/m_{I} \right|$, and if the conserved angular momenta $L_{I}$ for the particles $I=(1)$ and $(2)$ are negative, the center-of-mass energy of a head-on collision of two particles at the wormhole throat~$\rho=0$ is given by
\begin{eqnarray}
E_{CM}^{2}(0)
&\sim& \frac{8a^{2}L_{(1)}L_{(2)}}{b^{6}}+\frac{8a^{2}\left| L_{(1)} \right| \left| L_{(2)} \right|}{b^{6}}\nonumber\\
&=& \frac{16a^{2}L_{(1)}L_{(2)}}{b^{6}}.
\end{eqnarray} 
$E_{CM}^{2}(0)$ can be very large even if the initial energy at infinity is small. We thus find that rotating Teo wormholes can be particle accelerators if they are small.

\section{Summary and Discussion}
In the Teo wormhole spacetime, particles can reach the wormhole throat from an initial point in the flat regions in a finite proper time. 
This is an important difference with respect to the BSW process in an extremal Kerr black hole spacetime, 
because here a collision with a high center-of-mass energy is possible in a finite proper time. 
Moreover, we do not require any fine-tuning of the particle angular momentum or of the particle charge 
in order to reach a high center-of-mass energy.

The center-of-mass energy of the particles near the wormhole throat can be very large 
if the wormhole is rotating fast and if its size is small, namely the parameter 
$b$ is small ($b\ll a^{1/2}$, $\left| L_{I}/E_{I} \right|$, and $\left| L_{I}/m_{I} \right|$). 
If the energy of the center-of-mass is sufficiently high, the collision could generate a black hole near the wormhole. 
Our result can thus be interpreted as an indication that fast-rotating wormholes with small $b$ are unstable 
against particle collisions in the vicinity of the wormhole throat. 
We note that such a conclusion is independent of the matter content responsible for the existence of the wormhole. 
Usual considerations on the wormhole stability are instead based on the study of small perturbations around the background metric 
and the matter configuration. 
Because of this dependence of the matter content, 
in the wormhole cases it is not really possible to point out any analogy 
between particle collisions with high center-of-mass energy and instabilities depending on the matter content. 
However, the presence of an instability related to the collision of two particles 
with a high center-of-mass energy might become a useful tool to investigate the stability of gravitating objects.

Even if we accept the hypothesis that some observed astrophysical objects like the supermassive bodies in galactic nuclei might be wormholes, 
the high rotating condition would not be achieved. 
Our result may instead be relevant for wormhole formations.
The creation of wormholes is still an open problem, although some authors are challenging this issue (see e.g.~\cite{Maeda:2008bh} and the references therein).
Wormholes might have been created in the early Universe, because of quantum fluctuations in spacetime topology~\cite{Wheeler:1957mu},
and enlarged to classical size~\cite{Roman:1992xj}.
If the particle collision with high center-of-mass energy in rotating wormhole spacetimes implies the instability of a wormhole, 
we may constrain the initial condition of rotating wormholes and the growth process.

The fact that the collision of two particles has a high center-of-mass energy does not imply that an observer at infinity can detect very high energy radiation produced in the process, as such a radiation may be strongly affected by gravitational redshift. For instance, very energetic particles created in a collisional Penrose process~\cite{Piran_Shaham_Katz_1975} in an extremal or almost extremal Kerr black spacetime cannot escape to infinity with a high energy or equivalently, very massive particles cannot reach a distant observer~\cite{Bejger:2012yb, Zaslavskii:2012yp, Harada:2012ap}. The result is different in the Reissner-Nordstr\"om black hole case, in which a charged particle with a very high energy or a very large rest mass can do it~\cite{Zaslavskii:2010aw, Zaslavskii:2012ax, Nemoto:2012cq}. Recently, the details of the collisional Penrose process in an extremal Kerr spacetime have been re-examined~\cite{newref2}.

Zaslavskii investigated a head-on particle collision and a collisional Penrose process in a general spacetime without horizons or naked singularities~\cite{Zaslavskii:2013nra}. Using the notation in our Eq.~(\ref{eq:Metric}), he considered a spacetime with $N > 0$ but arbitrary small at the collision point. In this sense, the spacetime is almost extremal, although there is no horizon. The Teo wormhole spacetime discussed in this paper is quite different. As it can be seen from Eq.~(\ref{eq:N}), $N$ is always larger than 1, so it never becomes small. Our process is thus the first example of particle collision with high center-of-mass energy in a spacetime with no event horizons, no naked singularities, and not being extremal at all.

A rapidly-rotating Teo wormhole has an ergoregion where a collisional Penrose process can occur like in the ergosphere in the Kerr spacetime and in the generalized ergoregion~\cite{Denardo_Ruffini_1973,Denardo_Hively_Ruffini_1974} in the the Reissner-Nordstr\"om spacetime.
The details of the collisional Penrose process in the Teo wormhole spacetime will be discussed in a future work.

Our result does not mean that any static and spherical symmetric wormhole cannot accelerate particles. We can consider static and spherical symmetric wormhole spacetimes in which particles with small conserved energy can reach the wormhole throat and therefore some wormholes can be used to accelerate particles, see e.g. Ref.~\cite{new-ref}, pages 159-160.

We conclude with a comment on exact solutions of rotating wormholes. Recently, rotating Ellis wormholes with a phantom scalar field were investigated numerically in four~\cite{Kleihaus:2014dla} and in five dimensions~\cite{Dzhunushaliev:2013jja}. The authors found that there is an upper bound for the wormhole angular momentum. However, this does not imply that highly rotating wormholes cannot exist, since the key point is the matter content responsible for the spacetime geometry. The Teo wormhole metric was introduced to study concretely the properties of wormholes that can rotate fast and do not have any horizon~\cite{Teo:1998dp}. It is not an exact solution generated by a specific matter content. We hope that this paper will stimulate the investigations of rapidly-rotating wormholes.

\section*{Acknowledgements}
The authors would like to thank Tomohiro Harada, Ken-ichi Nakao, and Oleg B. Zaslavskii
for valuable comments.
The authors were supported by
the NSFC grant No.~11305038, the Shanghai Municipal
Education Commission grant for Innovative Programs
No.~14ZZ001, the Thousand Young Talents Program, and
Fudan University.

\appendix

\section{Particle collision in the Ellis wormhole spacetime}

This appendix is devoted to the collision of two particles in the Ellis wormhole spacetime.
The Ellis wormhole~\cite{Ellis_1973} was the first example of traversable wormhole of the Morris-Thorne class~\cite{Morris_Thorne_1988}. The line element is given by
\begin{eqnarray}
ds^{2}=-dt^{2}+d\rho^{2}+(\rho^{2}+b^{2})(d\theta^{2}+\sin^{2}\theta d\phi^{2}),
\end{eqnarray}
where $b$ is a positive constant. Since the spacetime is spherical symmetric, we can consider the motion on the equatorial plane $\theta=\pi/2$ without loss of generality. As we have done in the Teo wormhole case, we can define an effective potential $V_{eff}(\rho)$ as
\begin{eqnarray}
V_{eff}(\rho)\equiv \frac{1}{2} \left( m^{2}-E^{2}+\frac{L^{2}}{\rho^{2}+b^{2}} \right).
\end{eqnarray}
The effective potential $V_{eff}(\rho)$ is a monotonically decreasing function of $\rho^2$ and has a maximum at the throat $\rho=0$.

The center-of-mass energy of the collision of two particles is
\begin{eqnarray}
E_{CM}^{2}
&=&m_{(1)}^{2}+m_{(2)}^{2}+2E_{(1)}E_{(2)}-2\frac{L_{(1)}L_{(2)}}{\rho^{2}+b^{2}}\nonumber\\
&&-2\sigma_{\rho(1)}\sigma_{\rho(2)}\sqrt{R_{(1)}(\rho)}\sqrt{R_{(2)}(\rho)},
\end{eqnarray} 
where 
\begin{eqnarray}
R_{I}(\rho)\equiv-m_{I}^{2}+E_{I}^{2}-\frac{L_{I}^{2}}{\rho^{2}+b^{2}}
\end{eqnarray} 
for $I=(1)$ and $(2)$. At the throat $\rho=0$, the center-of-mass energy in a head-on collision of two particles and the effective potential are exactly the same as in the Teo wormhole spacetime with $a=0$. Therefore the Ellis wormhole cannot be a high-energy particle collider, just like the static and spherical symmetric Teo wormhole.

\section{Scalar Curvature}

The scalar curvature $\mathcal{R}$ on the equatorial plane ($\theta=\pi/2$) in the Teo wormhole is given by
\begin{eqnarray}\label{eq:scalar_curvature}
\mathcal{R}=-\frac{2 a^{2} \left( 64r^{4}(\rho)d+9b-9r(\rho) \right)}{r^{7}(\rho)}.
\end{eqnarray} 
As shown in Sec.~III, in the case of a wormhole of small size and high spin, namely $b\ll a^{1/2}$, $\left| L_{I}/E_{I} \right|$, and $\left| L_{I}/m_{I} \right|$, the Teo wormhole can work as a high-energy particle collider. It turns out that, in such a situation, the absolute value of the scalar curvature $\mathcal{R}$ is very large at the throat on the equatorial plane. On the contrary, in the non-rotating case $a=0$, the Teo wormhole cannot be used as a high-energy particle collider and Eq.~(\ref{eq:scalar_curvature}) shows that the scalar curvature $\mathcal{R}$ vanishes everywhere.

Such a finding between curvature and center-of-mass energy may suggest that the possibility of an efficient particle acceleration requires a large $|\mathcal{R}|$. Actually, this is not true. The scalar curvature $\mathcal{R}$ in the Ellis wormhole is given by
\begin{eqnarray}
\mathcal{R}=-\frac{2 b^{2}}{ \left( b^{2}+\rho^{2} \right)^2}.
\end{eqnarray} 
At the wormhole throat $\rho=0$, this gives $\mathcal{R}=-2/b^{2}$. In the small $b$ case, the absolute value of the scalar curvature is large at the throat $\rho=0$. However, the Ellis wormhole with small $b$ cannot be a high-energy particle collider as seen in appendix~A. We thus conclude that the scalar curvature $\mathcal{R}$ is not a good indicator to understand the possibility of the creation of high energy collision between two particles.

\end{document}